\documentclass[twocolumn]{aastex63}
\usepackage{bm}
\usepackage{soul}
\usepackage{xfrac}
\usepackage{xcolor}
\usepackage{amsmath}
\usepackage{CJKutf8}
\usepackage{hyperref}
\usepackage{textcomp}
\usepackage{multirow}
\usepackage{graphicx}
\usepackage{booktabs}
\usepackage{tabularx}
\usepackage{rotating}
\usepackage{makecell}
\usepackage{mathrsfs}
\usepackage{ragged2e}
\usepackage[T1]{fontenc}
\usepackage[utf8]{inputenc}

\newcommand{\cs}{c_{\rm s}}
\newcommand{\rhog}{\rho_{\rm g}}
\newcommand{\sigmag}{\Sigma_{\rm g}}
\newcommand{\taumean}{\langle\tau\rangle_\phi}
\newcommand{\tauprime}{\langle\tau^\prime\rangle}
\newcommand{\tauclump}{\langle\tau\rangle_{\rm clump}}
\newcommand{\simpchinese}[1]{\begin{CJK}{UTF8}{gbsn}#1\end{CJK}}
\newcommand{\tradchinese}[1]{\begin{CJK}{UTF8}{bsmi}#1\end{CJK}}

\begin{document}

\title{Dust Dynamics in Transitional Disks: Clumping and Disk Recession}

\author[0000-0002-0605-4961]{Jiaqing Bi \simpchinese{(毕嘉擎)}}
\affiliation{Department of Physics \& Astronomy, University of Victoria, 3800 Finnerty Road, Victoria, BC V8P 5C2, Canada; \url{bijiaqing@uvic.ca}}
\affiliation{Department of Astronomy, University of California at Berkeley, Campbell Hall, Berkeley, CA 94720, USA}
\affiliation{School of Physics, Xi'an Jiaotong University, 28 Xianning West Road, Xi'an, Shaanxi 710049, China}

\author[0000-0002-7455-6242]{Jeffrey Fung \tradchinese{(馮澤之)}}
\affiliation{Department of Physics \& Astronomy, Clemson University, 118 Kinard Laboratory, Clemson, SC 29634, USA; \url{fung@clemson.edu}}
\affiliation{Institute for Advanced Study, 1 Einstein Drive, Princeton, NJ 08540, USA}
\affiliation{Department of Astronomy, University of California at Berkeley, Campbell Hall, Berkeley, CA 94720, USA}


\begin{abstract}
    
    The role of radiation pressure in dust migration and the opening of inner cavities in transitional disks is revisited in this paper. Dust dynamics including radiation pressure is often studied in axisymmetric models, but in this work, we show that highly non-axisymmetric features can arise from an instability at the inner disk edge. Dust grains clump into high density features there, allowing radiation to leak around them and penetrate deeper into the disk, changing the course of dust migration. Our proof-of-concept, two-dimensional, vertically-averaged simulations show that the combination of radiation pressure, shadowing, and gas drag can produce a net outward migration, or recession, of the dust component of the disk. The recession speed of the inner disk edge is on the order of $10^{-5}$ times Keplerian speed in our parameter space, which is faster than the background viscous flow, assuming a \citeauthor{shakura_reprint_1973} viscosity $\alpha \lesssim 10^{-3}$. This speed, if sustained over the lifetime of the disk, can result in a dust cavity as large as tens of au.

\end{abstract}

\keywords{Protoplanetary disks (1300); Planet formation (1241); Circumstellar dust (236); Astrophysical dust processes (99); Astronomical simulations (1857)}


\section{Introduction}

Transitional disks are characterized by the depletion of dust in their central regions. However, the origin of those inner cavities is still uncertain. Recent surveys on transitional disks by Atacama Large Millimeter/Submillimeter Array (ALMA) and Australia Telescope Compact Array (ATCA) suggest that the cavities are cleared by planets in the disk (e.g., \citealt{van_der_marel_new_2018, francis_dust-depleted_2020, norfolk_dust_2021}). However, evidence shows that planets may not be the only explanation for these cavities:
\begin{enumerate}
    \item There are only two confirmed planets in transitional disks so far (PDS 70 b/c; \citealt{keppler_discovery_2018, haffert_two_2019}), even though the dust-depleted, optically thin cavities should provide a good chance to detect planets via high-contrast direct imaging.
    \item A complete disk survey in the Lupus star-forming region shows that the fraction of disks with a large cavity ($\geq$20 au) is $\gtrsim$11\% \citep{van_der_marel_new_2018}. This fraction is noticeably higher than the fraction of giant planets on such high orbits around main sequence stars (<8\% for all stars, <6.8\% for FGK stars, and <4.2\% for M stars; \citealt{bowler_imaging_2016}).
    \item The transitional disk fraction for Herbig Ae/Be stars ($\sim$28\%; \citealt{guzman-diaz_homogeneous_2021}) is typically higher than that for T Tauri stars ($\lesssim$10\%; \citealt{muzerolle_spitzer_2010, furlan_spitzer_2011}).
\end{enumerate}

Therefore, we want to seek for other possible mechanisms to clear the cavity in transitional disks. Since the cavities appear to be correlated with the spectral type of the central star, mechanisms related to the stellar luminosity such as photoevaporation and radiation pressure may be responsible for cavity clearing. However, the Herbig Ae/Be star survey cited above \citep{guzman-diaz_homogeneous_2021} reveals that some transitional disks have ages less than
0.1 Myr. This is shorter than the estimated timescale for photoevaporation to open the gap ($\gtrsim$1 Myr; \citealt{garate_large_2021}). Instead, we revisit an alternative explanation: dust migration due to radiation pressure.

Radiation pressure from the central star pushes dust grains outward, but whether it can prevent dust from accreting onto the star is still under debate. \cite{takeuchi_dust_2001} studied the dynamics of irradiated dust grains in an optically thin, non-accretion disk. They pointed out that the angular momentum of dust grains can be modified by radiation pressure, such that dust grains can migrate inward/outward when they move faster/slower than the gas azimuthally. \cite{takeuchi_surface_2003} then extended their study to optically thick accretion disks. They found that in a disk similar to the minimum mass solar nebula (MMSN), dust grains only migrate outward on the disk surface, and the outward flux is negligible compared with the inward accretion flow on the midplane. \cite{chiang_inside-out_2007} suggested that radiation pressure can blow out dust grains in a gaseous rim that is unstable to the magneto-rotational instability. Later, \cite{dominik_accretion_2011} argued that it cannot. They showed that in an one-dimensional (1D), axisymmetric, vertically averaged picture, radiation pressure can be overcome by the combined effects of dust pile-up and gas accretion. Then, \cite{owen_radiation_2019} added grain size evolution to the analysis, and generalized it to 2D (radial and vertical). They showed that at a pressure bump created by photoevaporation, small dust grains are lifted to higher altitudes, where they are removed by radiation pressure and then replenished by the fragmentation of larger grains, ultimately leading to the removal of all dust. In this picture, the transitional disk cavity would be cleared by photoevaporation (e.g., \citealt{picogna_dispersal_2019, ercolano_dispersal_2021}). \cite{krumholz_dynamics_2020} revisited the problem for smooth disks without pressure bumps, and showed that for low-mass disks ($\lesssim$1\% of MMSN) with accretion rates lower than $\sim$10$^{-11} M_\odot$ per year, radiation pressure can remove micron-sized grains from the inner disk even without the aid of a pressure bump. So it seems, radiation pressure cannot be responsible for transitional disk cavities if the disks were as gas-rich as MMSN.

In this paper, we show that there is another level of complexity that previous studies have not considered, and it can open an avenue to cavity-clearing even for gas-rich disks. In essence, an instability at the disk edge can break the axisymmetry that was commonly assumed, generating a host of rich and new dust dynamics, including the formation of small, high density dust features that we will refer to as \textit{clumps}. We will demonstrate this directly using 2D, vertically averaged numerical models. The instability we will describe is similar to the irradiation instability (IRI) in \cite{fung_irradiation_2014} as both of them are associated with radiation and extinction, and require a sharp radial transition between optically thin and thick. However, we will show that the instability is different from IRI as it does not require the participation of the gas component (except for the drag force on the dust).

The paper is organized as follows: We first describe our methodology of simulating dust grains and the numerical modeling in Section \ref{sec:algorithm}. We present our results in Section \ref{sec:results}, starting with the clumps at the asymmetric disk edge, and followed by the disk edge recession. We discuss the possible effects of a few neglected mechanisms in Section \ref{sec:discussion}. Finally, we conclude in Section \ref{sec:conclusion}.

\section{Numerical Method} \label{sec:algorithm}

We develop a graphics processing unit (GPU) particle code to simulate dust grains in a gaseous disk. It employs the staggered semi-analytic method \citep[SSA;][]{fung_staggered_2019}, which is designed to achieve high accuracy when the equation of motion is ``stiff'', i.e., when the stopping time of dust grains is much shorter than the integration time step, which is the case for our models, where the Stokes number can be as small as $10^{-5}$.

Dust grains in our models interact with each other through shadowing. We compute the optical depth of the dust disk in 2D grids using the ``cloud-in-cell'' (CIC) prescription (e.g., Chapter 1.5.2, \citealt{hockney_computer_1981}). Each simulated particle, called ``super-particle'', represents a cloud of dust grains having the size of one grid cell. To compute the optical depth, we first evaluate the disk's dust surface density $\Sigma$ by distributing the mass of each super-particle into its four closest neighboring grid cells through bilinear interpolation. The $\Sigma$ grids are then converted into the $\tau$ grids using the opacity of dust $\kappa_{{\rm opa}}$:
\begin{equation} \label{eq:tau}
\tau(r, \phi) = \int^{r}_{0} \kappa_{{\rm opa}}\,\frac{\Sigma(r^\prime,
\phi)}{h(r^\prime)}\,{\rm d}r^{\prime} \, ,
\end{equation}
where $h$ is the disk scale height, which we assume to be proportional to the radius $r$ such that the disk aspect ratio $h/r$ is a constant. The values of $\kappa_{{\rm opa}}$ and $h/r$ will be discussed in the following section. Finally, the $\tau$ value at the position of a given super-particle is extracted from the $\tau$ grids again through bilinear interpolation.

The CIC method of computing $\tau$ estimates the averaged $\tau$ in the cloud of dust grains that a single super-particle represents. This implies that super-particles can self-shadow; in other words, even when there is no dust between the star and a super-particle, the $\tau$ value at its position will still not be zero because the front end of the cloud casts a shadow on its back end. The amount of self-shadowing is small, equaling about a half of the optical thickness of the super-particle, and is generally not consequential, but it does have an effect when there exist an optically thick concentration of dust on a scale smaller than a grid cell.


\subsection{Model Description} \label{sec:model}

We consider a 2D protoplanetary disk where dust grains and a steady gaseous disk orbit a central star of mass $M_\star$. We use $\{r, \,\phi \}$ to denote radius and azimuth in the 2D polar coordinates, respectively. The coordinate system is fixed and centered on the star. Hereafter, we use the subscript `0' to denote any evaluations at the reference radius $r_0$.

We use $10^7$ super-particles in our fiducial model to resolve our dust disk. Each of them represents a cloud of dust grains that share the same internal density $\rho_\bullet$ and grain size $s$. In our models, all super-particles have the same $s$. We will discuss possible effects of the grain size distribution in Section \ref{sec:collision}.

To compute gas drag on the dust, we assume the grains are subjected to the sub-sonic Epstein drag, such that the stopping time is $t_{\rm s} = \rho_\bullet s/\rhog\cs$ \citep{weidenschilling_aerodynamics_1977}, where $\rhog$ and $\cs$ are the volumetric density and sound speed of gas, respectively. The parametrized evaluation of $t_{\rm s}$ reads:
\begin{equation} \label{eq:tstop}
t_{\rm s} = \frac{\Sigma_{\rm g0}c_{\rm s0}}{\sigmag \cs} \frac{{\rm St}_0}{\Omega_{\rm K0}},
\end{equation}
where $\sigmag$ is the surface density of gas, ${\rm St}$ is the Stokes number, and $\Omega_{\rm K}(r) = \sqrt{GM_\star/r^3}$ is the Keplerian angular velocity. Epstein drag applies when the grain size is insignificant compared with the mean free path of gas, which is typically satisfied for micron-size grains in protoplanetary disks. In our parameter space, we also expect that gas drag can keep the dust-to-gas relative motion subsonic; retrospectively, indeed we find in our models that the dust-to-gas speed is typically slower than 1\% of the sound speed.

Given this drag force, we integrate the following equation of motion in our code:
\begin{equation}
    \bm{a} = \bm{f}(t,\bm{x},\bm{v}) 
           + \frac{\bm{v}_{\rm g}(t,\bm{x}) - \bm{v}}{t_{\rm s}(t,\bm{x})},
\end{equation}
where $\bm{a}, \bm{x}, \bm{v}$ are the acceleration, position, and velocity of dust, $\bm{f}$ is the specific force acting on the dust except for gas drag, and $\bm{v}_{\rm g}$ is the gas velocity. In this study, the radial and azimuthal components of $\bm{f}$ are
\begin{align}
f_{\rm r} & = -\frac{GM_\star}{r^2}(1 - \beta e^{-\tau}), \\
f_\phi & = 0.
\end{align}
The absolute ratio between radiation pressure and gravity is
\begin{equation} \label{eq:beta}
\beta = \frac{\kappa_{\rm abs}L}{4\pi c\,GM_\star},
\end{equation}
where $\kappa_{\rm abs}$ is the particle mass absorption coefficient, \textit{L} is the stellar luminosity, and \textit{c} is the speed of light. Micron-sized dust grains have typical $\beta$ values of $\sim$0.2--4 for T Tauri stars and $\sim$4--4000 for Herbig Ae/Be stars \citep{garufi_disks_2020, guzman-diaz_homogeneous_2021}. We note that $\kappa_{\rm abs}$ is different from $\kappa_{\rm opa}$ used in Equation \ref{eq:tau}, as the opacity $\kappa_{\rm opa}$ includes both extinction due to absorption ($\kappa_{\rm abs}$) and scattering. In our models, we do not explicitly set the value of all parameters on the right-hand-side of Equation \ref{eq:beta}. Instead, we set values for $\beta$, which allows us to generalize our models to any combination of $L$, $M_\star$, and $\kappa_{\rm abs}$.


\subsection{Initialization} \label{sec:init}

We assign the initial radial position of super-particles using a probability distribution of the total dust mass $M$ in the disk:
\begin{equation}
    \frac{{\rm d}M}{{\rm d}r}(r) = \left(2\pi r\Sigma^\prime \right)*G = 2\pi \int_0^\infty r\Sigma^\prime (r^\prime)\,G(r - r^\prime)\,{\rm d}r^{\prime},
\end{equation}
where $*$ is the convolution operator,
\begin{align}
\Sigma^\prime (r) & =
  \begin{cases}
  \Sigma_0 \left(r/r_0\right)^{-3/2}, & 1.0 \leq r/r_0 \leq 1.5; \\
  0,              & {\rm else,}
\end{cases} \\
G(r) & = \frac{1}{0.02 r_0 \sqrt{2\pi}}\exp\left({-\frac{r^2}{2(0.02 r_0 )^2}}\right),
\end{align}
$r_0$ is a constant that equals to 1 in code units, and $\Sigma_0$ 
is a normalization that we adjust according to the total optical depth of the disk, which we will describe in the following paragraphs. The initial surface density profile is therefore $\Sigma(r)=({\rm d}M/{\rm d}r)/(2\pi r)$, which is a smoothed-edge dust ring between roughly $1.0 r_0$ and $1.5 r_0$. The initial azimuthal position of the super-particles are assigned using a uniform probability distribution between $0$ and $2\pi$.

In this paper, we use $\langle\,\rangle_\phi$ to denote azimuthal averaging, such that $\taumean$ is the azimuthally averaged $\tau$. We also define 
\begin{equation} \label{eq:tauprime}
    \tauprime = -\log \langle e^{-\tau} \rangle_\phi
\end{equation}
to be the ``effective optical depth'' that relates to the azimuthally averaged amount of extinction. We refer the radius where $\tauprime = 1$ as the ``optical edge'' of the disk. Note that $\tau$, $\taumean$, and $\tauprime$ are only equal when the dust distribution is axisymmetric; they are not the same when the disk develops asymmetric structures. Later in Section \ref{sec:results}, we will be using their differences to quantify asymmetry in our disks.

The total amount of dust in our model is normalized according to $\tau$ of the disk. In our fiducial model, we set $\taumean$ at the peak of the $\Sigma$ profile ($r\sim1.05r_0$) to 10. To apply this normalization to $\Sigma$, we assume that $r_0$ = 1 au, $s$ = 1 $\mu$m, $\rho_\bullet$ = 1.5 g/cm$^{3}$, $h/r = 0.05$, and $\kappa_{\rm opa}$ is given by the geometric opacity:
\begin{equation} \label{eq:kappa}
    \kappa_{\rm opa} = \frac{\pi s^2}{\sfrac{4}{3}\,\pi s^3 \rho_\bullet} = 0.5 \left( \frac{s}{1 \,{\rm cm}} \right )^{-1} \,{\rm cm}^2{\rm g}^{-1}.
\end{equation}
It gives $\Sigma_0$ = $7.7\times10^{-3}$ g/cm$^2$ for micron-sized dust grains in our fiducial model\footnote{Hereafter, the unit of $\Sigma_0$ is g/cm$^2$, if unspecified.}. 
We emphasize that $\Sigma_0$ represents only the micron-sized grains --- if we assume that the dust grains follows a grain size distribution ${\rm d}n_{\rm s} \propto s^{-3.5} \,{\rm d}s$ \citep{mathis_size_1977} from 1 $\mu$m to 1 km, our model would produce a total dust density of about $2.4\times10^2$ g/cm$^2$, corresponding to a dust-to-gas ratio of about $0.14$, assuming the gas follows the minimum mass Solar Nebula at 1 au \citep{weidenschilling_distribution_1977, hayashi_structure_1981}. We will study the effects of different levels of $\tau$ by varying the value of $\Sigma_0$. 

For the purpose of computing gas drag, we assume a steady background gas surface density profile that follows $\sigmag \propto (r/r_0)^{-p}$ and a sound speed profile $\cs \propto (r/r_0)^{-q/2}$, where $p = 1.5$ and $q = 1$. The gas velocities assume hydrostatic equilibrium:
\begin{align}
v_{r, \,{\rm gas}} & = 0, \\
v_{\phi, \,{\rm gas}} & = v_{\rm K} \sqrt{1 - \eta}, 
\end{align}
where $v_{\rm K} = r \Omega_{\rm K}$ is the Keplerian velocity, and $\eta = (p + q)(h/r)^2$ is the pressure-related term. The gas density is not explicitly normalized, but rather incorporated into the value of the Stokes number ${\rm St}_0$.

When initializing our super-particles, we assign them Keplerian orbital velocities:
\begin{align}
v_r & = 0, \\
v_\phi & = v_{\rm K}.
\end{align}
As soon as the simulation begins, gas drag will cause a headwind on the super-particles and lead to their radial drift; at the same time, radiation pressure will push the super-particles outward if they lie in the optically thin disk edge. Although these motions are not captured in the initialization, the values of ${\rm St}_0$ we use are sufficiently small (i.e., the gas-dust coupling is sufficiently strong) such that the super-particles can adjust to gas drag within a dynamical time $\Omega_{\rm K}^{-1}$. Subsequently, the dust dynamics is not sensitive to the initial velocities.

Finally, for our parameter space study, we consider $\beta$ values ranging from 0.01 to 20, ${\rm St}_0$ from $10^{-5}$ to $10^{-2}$, and $\Sigma_0$ from $2.3\times10^{-3}$ to $7.7\times10^{-2}$ g/cm$^2$ (i.e., $\tau$ from 3 to 100 at the radius of the initial surface density peak). We refer to the $\{\beta,\,{\rm St}_0,\,\Sigma_0\}=\{10,\,10^{-4},\,7.7\times10^{-3}\}$ setup as our fiducial model. 


\subsection{Domain and Resolution}

Our $\Sigma$ and $\tau$ grids are defined in a domain that extends radially from 0.95 to 1.55 $r_0$, and azimuthally around the full 2$\pi$. $\Sigma$ is defined at the center of each grid cell, and set to zero outside the domain. $\tau$ is defined at the center of the outer radial boundary of each grid cell, set to zero within the inner domain boundary, and infinity beyond the outer domain boundary. In other words, the disk is always optically thin within $0.95r_0$ and optically thick outside $1.55r_0$. In most of our models, super-particles do not leave the domain.

The fiducial resolutions are $N_{\rm r} \times N_\phi = 1024 \times 1024$, with grid cells spaced logarithmically in the radial direction and uniformly in the azimuthal direction. We adopt a fixed time step $\delta t = 2\pi / (N_\phi \Omega_{\rm K0})$ that is sufficient to ensure particles do not skip grid cells during integration. In the following section, we examine the validity of these choices by running convergence tests.


\begin{figure}[tp]

\centering
\includegraphics[width = 0.45\textwidth]{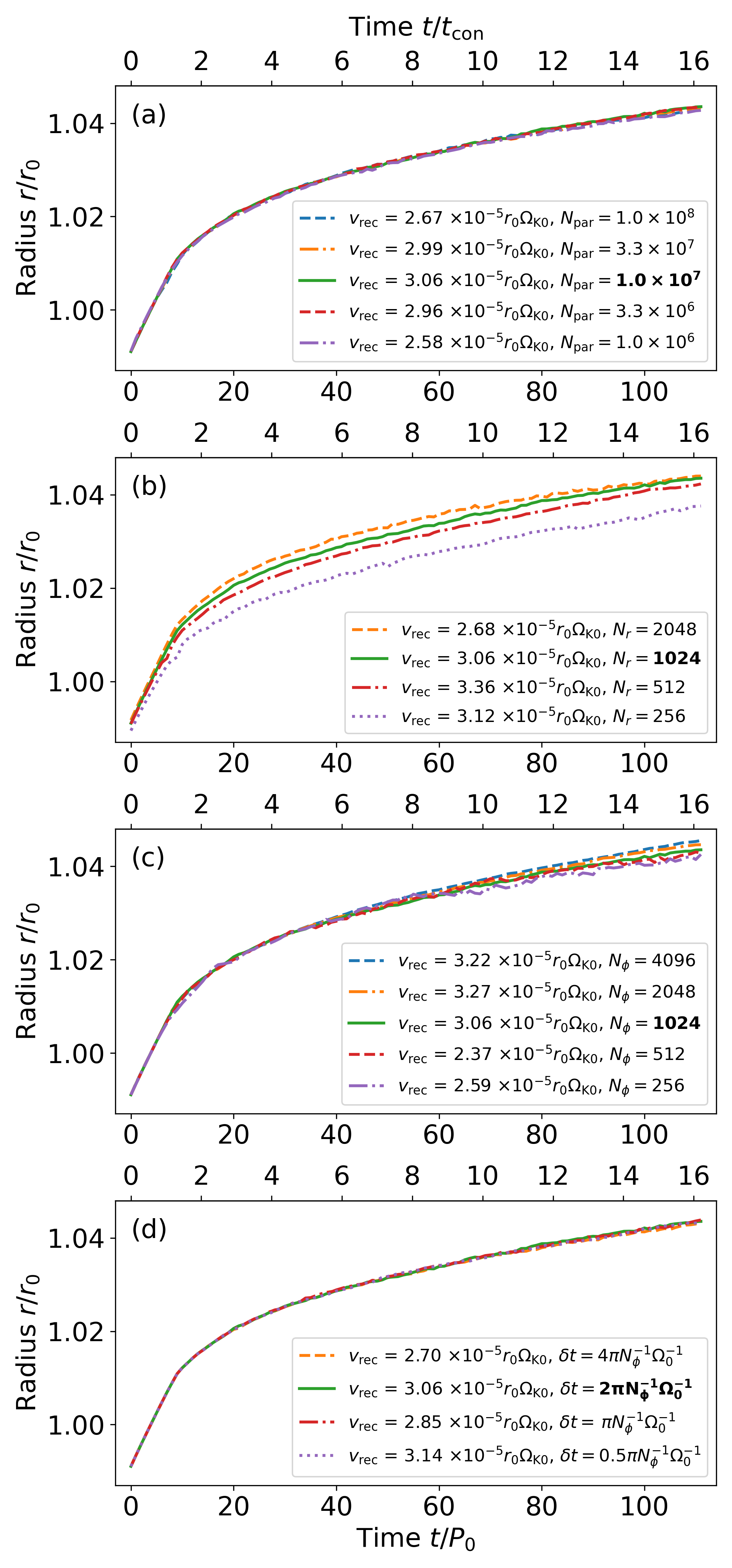}
\figcaption{The result of convergence tests using the time evolution of the optical edge (where $\tauprime \equiv 1$). \textbf{(a):} the convergence test of the super-particle resolution $N_{\rm par}$. \textbf{(b):} the convergence test of the radial resolution $N_r$. \textbf{(c):} the convergence test of the azimuthal resolution $N_\phi$. \textbf{(d):} the convergence test of the time step $\delta t$. The top axis labels the simulation time in the unit of concentration timescale (see Section \ref{sec:formation}). The bottom axis labels the simulation time in the unit of orbital period. The legend labels the recession speed of the optical edge (see Section \ref{sec:recession}), with the fiducial parameters bold marked.
\label{fig:numparamtest}}
\end{figure}

\subsection{Convergence Test of Numerical Parameters} \label{sec:numparamtest}

We run convergence tests for particle resolution $N_{\rm par}$, radial resolution $N_r$, azimuthal resolution $N_\phi$, and time step $\delta t$. The aim is to obtain converged results on the movement of the optical edge (where $\tauprime = 1$). All tests are performed using our fiducial set of $\{\beta,\,{\rm St}_0,\,\Sigma_0\}$. The results are shown in Figure \ref{fig:numparamtest} and described below.

It is important to note that the edge instability that drives the evolution of these models grows from the random noise in the initialization of the super-particles. As we vary numerical parameters, that initial noise is changed and that leads to a slightly different simulation, even if all physical parameters remain unchanged. This means we do not expect to find precise convergence, but we do consider our simulations robust as long as there is convergence in the phenomenological behavior of the models and that the quantitative measurements do not vary by more than $\sim10-20\%$.

\paragraph{Particle Resolution} We try super-particle numbers ranging from $10^6$ to $10^8$, and compare with the result of the fiducial value $10^7$. We find good convergence in the edge location as long as the number is larger than $10^6$, as seen in the top panel of Figure \ref{fig:numparamtest}.

\paragraph{Radial Resolution} We vary the radial grid resolution in this test. In the second panel of Figure \ref{fig:numparamtest}, we identify good convergence when $N_{\rm r}\gtrsim512$.

\paragraph{Azimuthal Resolution} Similar to the previous test, here we vary the azimuthal grid resolution. In the third panel of Figure \ref{fig:numparamtest}, we find good convergence in both the edge location and the recession speed within the time domain of the test simulations. Generally, the disks appear to recess faster with higher azimuthal resolutions. With higher azimuthal resolution, the size of each super-particle is effectively smaller (see Section \ref{sec:algorithm}), making it easier for radiation pressure to leak around opaque regions. So given the same initial distribution of super-particles, the true recession speed may be higher than what we measure, but unlikely to be lower.

\paragraph{Time Step} In our last test, we vary the time step. The last panel of Figure \ref{fig:numparamtest} shows that the edge location is well-converged.

\section{Results} \label{sec:results}

\begin{figure}[tp]
\centering
\includegraphics[width = 0.45\textwidth]{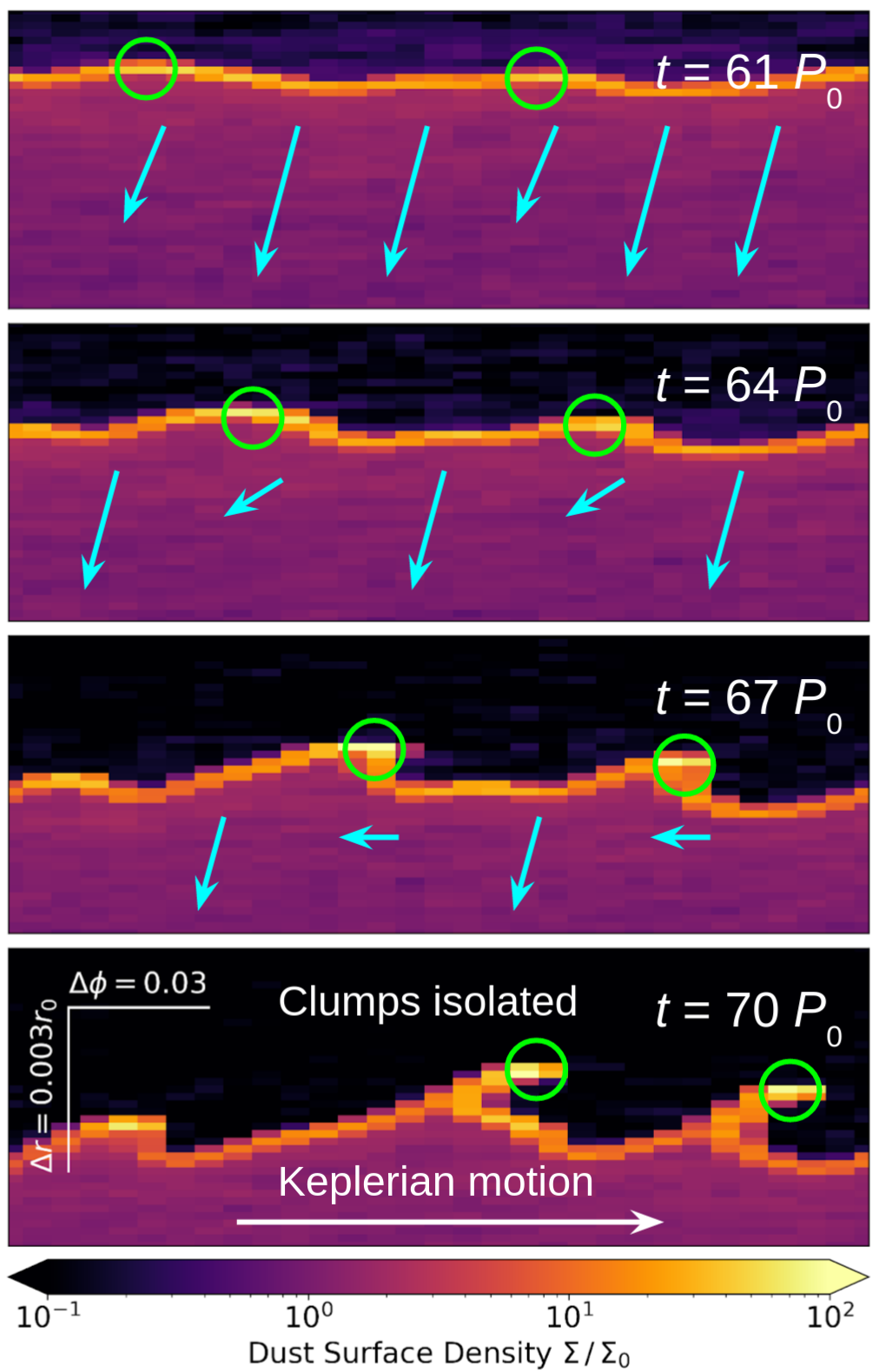}
\figcaption{Snapshots of dust surface density of the model with $\beta = 1$, ${\rm St}_0 = 10^{-4}$, and $\Sigma_0 = 7.7\times10^{-3}$ g/cm$^2$. The green circles mark the location of the overdensities at the disk edge, and the cyan arrows mark the motion of disk material at different azimuths relative to the overdensities. \label{fig:clump}}
\end{figure}

\subsection{The Asymmetric Disk Edge} \label{sec:clump}

Our simulations typically evolve as follows. In the initial phase, dust grains that are in the optically thin region are pushed outward by radiation pressure. This migration of dust leads to the formation of a sharp dust wall. Slight variations in density along this wall are amplified over time. The lower density segments are pushed back at higher speeds, and they become radially separated from the denser segments. The Keplerian shear then takes these lower density segments to the back of the high density segments. Due to shadowing, the front of the wall is always recessing faster than the back, this simulates a radial compression that adds to the density of the already dense segments, which we refer to as \textit{clumps}. This process is broken down in Figure \ref{fig:clump}. 

To create these asymmetric features, a sufficiently sharp dust wall must form first, in other words, $({\rm d}\tau/{\rm d}r)e^{-\tau}$ must be large, otherwise the Keplerian shear will smear the local density differences before the high and low density segments can separate radially. This is analogous to IRI, which also relies on the radial compression due to shadowing and requires a sharp transition from optically thin to thick.

\begin{figure}[tp]
\centering
\includegraphics[width = 0.45\textwidth]{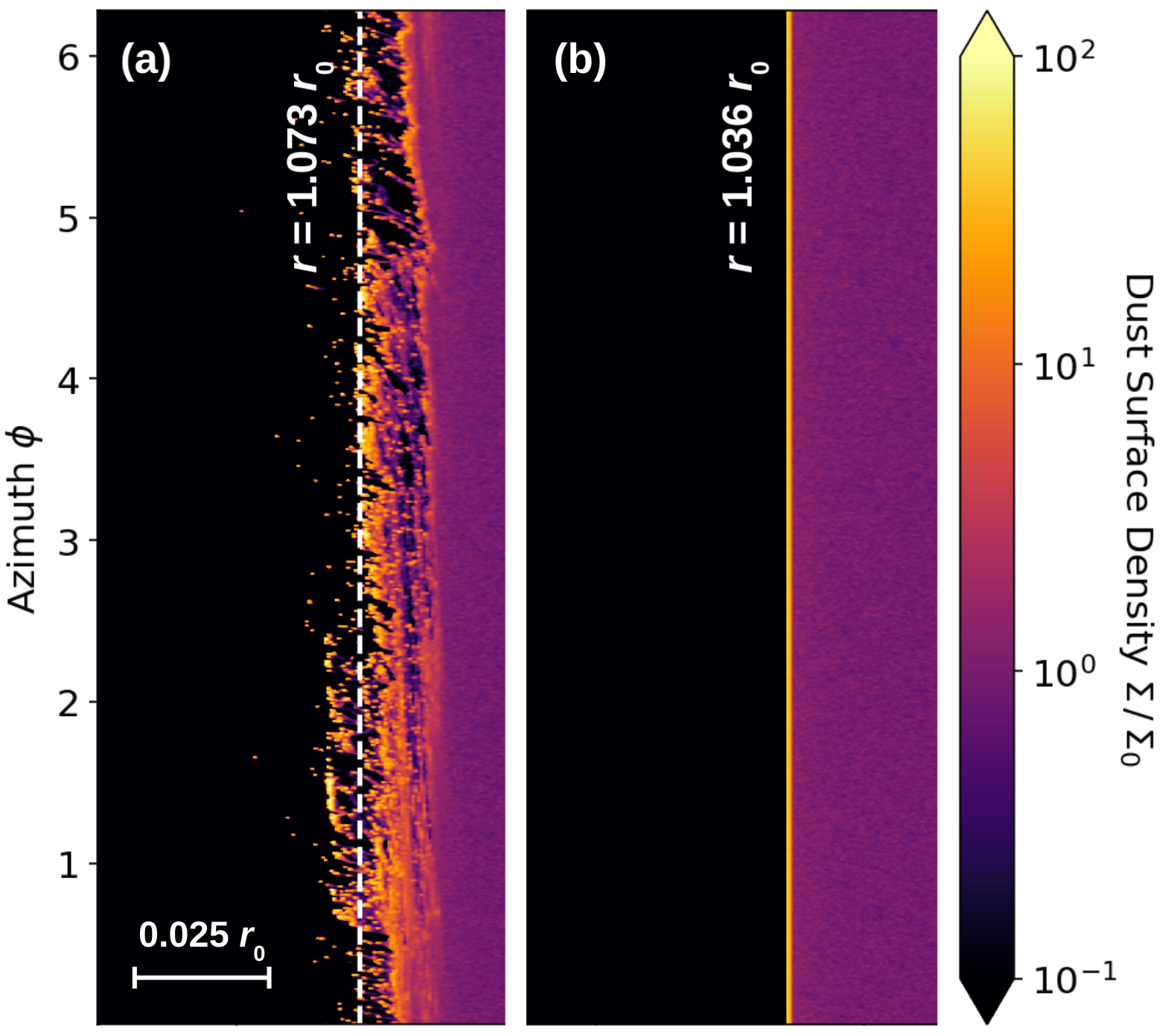}
\figcaption{Snapshots of dust surface density at $t = 500P_0$ in \textbf{(a)} the fiducial model, and \textbf{(b)} the instability inactive model. In the latter case, the optical depth $\tau$ is azimuthally averaged at every time step.
\label{fig:asym}}
\end{figure}

Common models of dust migration due to radiation pressure assume axisymmetry. Figure \ref{fig:asym} illustrates what would happen if we enforce such condition. On the right panel, we inactivate any instability by artificially setting $\tau$ equal to its azimuthally averaged value $\langle \tau\rangle_\phi$ at every timestep. The disk edge then develops into a dense but symmetric wall. We note that the disk edges are formed at different radii in the two models, which means that the clump formation has a qualitative effect on the radial migration of dust. We will discuss this in Section \ref{sec:recession}.


\subsubsection{Clump Formation and Destruction} \label{sec:formation}

\begin{figure}[tp]
\centering
\includegraphics[width = 0.45\textwidth]{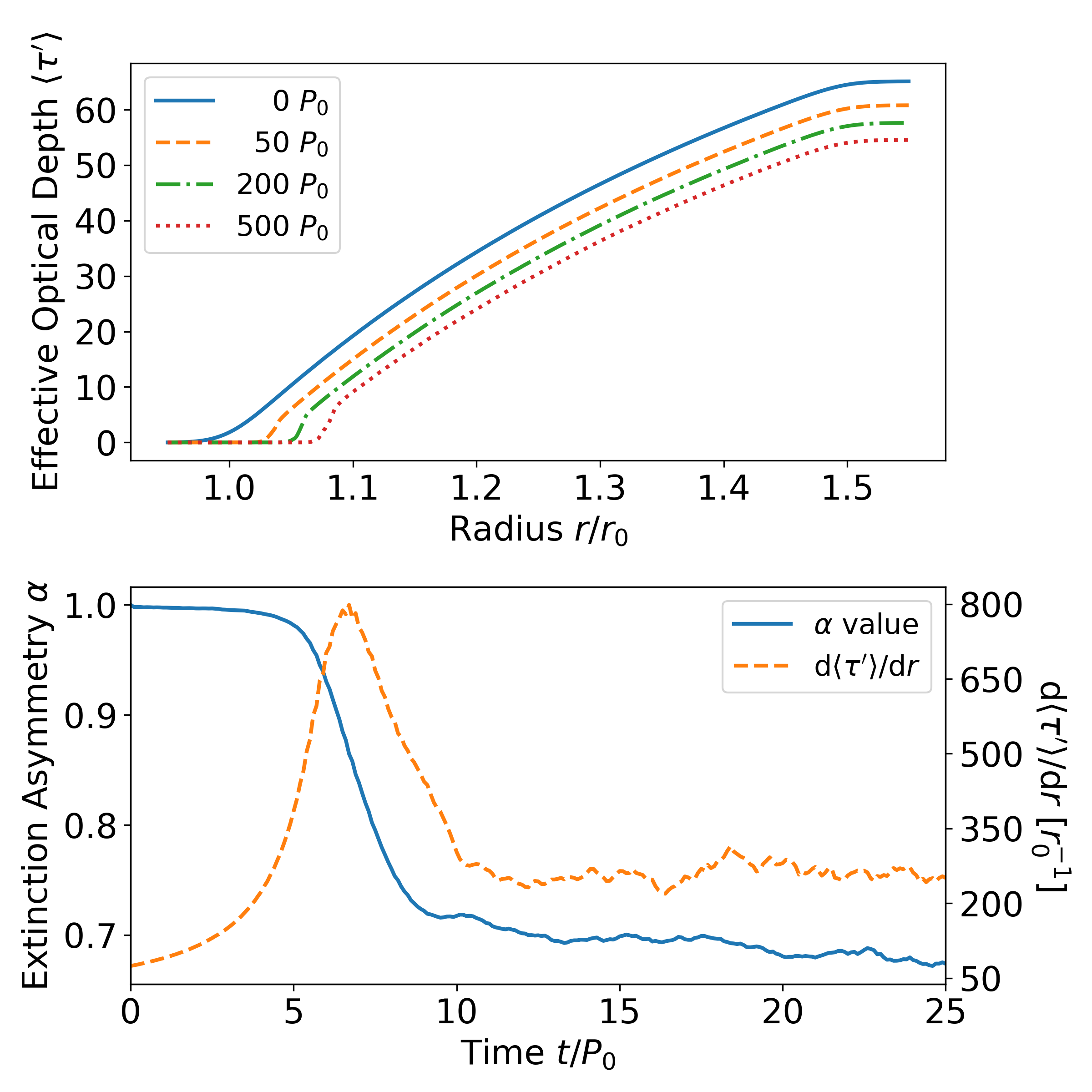}
\figcaption{\textbf{Upper:} the time evolution of the effective optical depth $\tauprime$ in the disk. \textbf{Lower:} the time evolution of the extinction asymmetry parameter $\alpha$ and the radial gradient of the effective optical depth ${\rm d}\tauprime/{\rm d}r$. Both values are measured at the optical edge (where $\tauprime \equiv 1$) in the fiducial model ($\beta = 10$, ${\rm St}_0 = 10^{-4}$, and $\Sigma_0 = 7.7\times10^{-3}$ g/cm$^2$).
\label{fig:alpha1}}
\end{figure}

The formation of high density dust clumps at the inner edge of illuminated disks is important in many ways. The one which we focus on in this paper is that they effectively reduce the optical depth of the disk (upper panel of Figure \ref{fig:alpha1}). This reduction allows radiation to penetrate deeper into the disk and change the course of dust migration. 

Instead of looking at individual clumps, we measure their collective influence using an ``extinction asymmetry'' parameter $\alpha$, which we define as:
\begin{equation}
    \alpha = \frac{1 - e^{-\tauprime}}{1 - e^{-\taumean}}.
\end{equation}
$\alpha$ is measured at the optical edge (where $\tauprime = 1$). Since $\tauprime < \taumean$ in general, we have $1-e^{-1} < \alpha \leq 1$ at the optical edge, with a smaller $\alpha$ denoting more effective clumping.

As mentioned, clump formation begins with the formation of a sharp dust wall. The lower panel of Figure \ref{fig:alpha1} shows the time evolution of $\alpha$ and the radial gradient of $\tauprime$ in the fiducial model. Initially, ${\rm d}\tauprime/{\rm d}r$ increases as the dust wall forms, while $\alpha$ remaining at $\sim$1. Once ${\rm d}\tauprime/{\rm d}r$ goes above $\sim$500$r_0^{-1}$, $\alpha$ starts to decrease, signaling clump formation. Isolated clumps, which are optically thick, are separated from the disk edge, which is still being pushed back by radiation pressure. This causes ${\rm d}\tauprime/{\rm d}r$ to decrease as $\alpha$ continues to fall. Eventually, both ${\rm d}\tauprime/{\rm d}r$ and $\alpha$ stabilizes. $\alpha$ stabilizes because it has almost approached the minimum value $1-e^{-1}$. ${\rm d}\tauprime/{\rm d}r$ stabilizes because the rates of clump formation and destruction are near equilibrium. 

In our simulations, clumps are not held together by any internal forces\footnote{Although, in reality they could be held together by self-gravity.}. When dust grains are apart from the clump body by roughly one grid size azimuthally, the exposure to the much stronger radiation pushes them outward rapidly. Therefore, the destruction timescale of clumps is related to our numerical resolution. Smaller clumps survive longer in our simulations, sometimes as long as our simulation time. The shredded clumps form a new wall behind the existing clumps, where the next generation of clump formation begins.

\begin{figure}[tp]
\centering
\includegraphics[width = 0.45\textwidth]{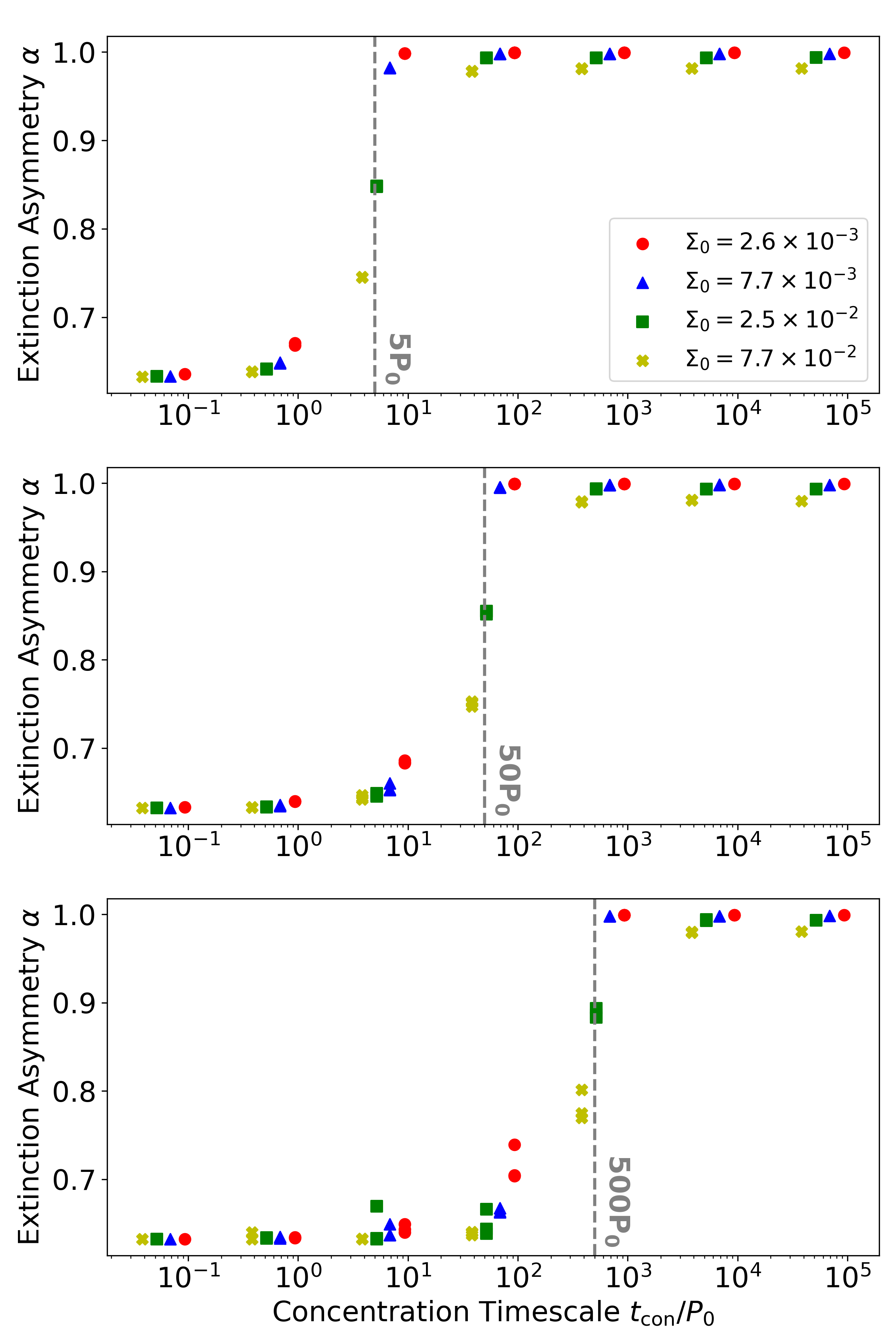}
\figcaption{The extinction asymmetry parameter $\alpha$ at the optical edge (where $\tauprime \equiv 1$) with different concentration timescale $t_{\rm con}$. Values are measured at $t = 5 P_0$ (\textit{upper}), $t = 50 P_0$ (\textit{middle}), and $t = 500 P_0$ (\textit{lower}). The parameter space for different $t_{\rm con}$ is listed in Table \ref{tab:paramspace}. We note that a large fraction of the sample is hidden by the overlapping of markers. Models with different $\Sigma_0$ suffer from different levels of Poisson noise that may affect the initial axisymmetry.
\label{fig:alpha2}}
\end{figure}

\begin{table}[tp]

\centering
\caption{The parameter space for studying $t_{\rm con}$} \label{tab:paramspace}

\medskip
\begin{tabular*}{0.45\textwidth}{@{\extracolsep{\fill}}cllll}

\toprule
$\beta$      & $1.0\times10^{-2}$ & $1.0\times10^{-1}$ & $1.0\times10^{0}$  & $1.0\times10^{1}$  \\ \\[-1.2em]
${\rm St_0}$ & $1.0\times10^{-5}$ & $1.0\times10^{-4}$ & $1.0\times10^{-3}$ & $1.0\times10^{-2}$ \\ \\[-1.2em]
$\Sigma_0$   & $2.6\times10^{-3}$ & $7.7\times10^{-3}$ & $2.5\times10^{-2}$ & $7.7\times10^{-2}$ \\ \\[-1.2em]

\bottomrule
\end{tabular*}

\justify
\tablecomments{We investigate all combinations of the listed values, such that there are 64 cases shown in each panel of Figure \ref{fig:alpha2}. The unit of $\Sigma_0$ is g/cm$^2$.}

\end{table}

There are a number of simplifications in our model that may affect the physics of clump formation. First, forming a sharp dust wall is a prerequisite for clump formation, but to form such a wall requires the radial concentration of the disk edge to overcome diffusive effects such as grain collisions and gas turbulence\footnote{We note that the lack of collision and turbulence in our study leads to the diffusion timescale being infinite, which means that the optical depth transition is always able to become sufficiently sharp to trigger effective clump formation.}. Here we define a concentration timescale
\begin{equation}
    t_{\rm con} = \frac{\Delta r_{\rm trans}}{\Delta v_{r,{\rm term}}}
\end{equation}
to estimate the time for the initial optical depth transition to become sufficiently sharp to trigger effective clump formation. $\Delta r_{\rm trans}$ is the radial extent of the optical depth transition that varies with $\Sigma_0$. In this study, we define $\Delta r_{\rm trans}$ numerically from where $\taumean = 0.1$ to where $\taumean = 1$ in the initial profile. $\Delta v_{r,{\rm term}}$ is the difference between the radial terminal velocity at the inner and outer edges of the optical depth transition discussed above (see Equation \ref{eq:terminal}). 

This concentration time $t_{\rm con}$ determines the time to form a dust wall, and therefore the time to form clumps. In Figure \ref{fig:alpha2} we show $\alpha$ values at the optical edge with a variety of $\beta$, ${\rm St}_0$, and $\Sigma_0$ (see Table \ref{tab:paramspace}) at different simulation time. It is clear that when $t = t_{\rm con}$, the initially axisymmetric optical edge starts to become clumpy, and ends up with $\alpha \sim 1-{\rm e}^{-1}$ when $t \gg t_{\rm con}$. The result remains self-similarity over time. Given sufficient time, all of our diffusion-less models will develop clumps. With some diffusion, clump formation may cease to occur if $t_{\rm con}$ is too long.

We also note that the parameter space in Table \ref{tab:paramspace} focuses on the regime where the gas-dust coupling is strong (${\rm St}_0 \ll 1$). This is because a certain level of gas drag is necessary to form clumps. Without the dissipation from gas drag, dust grains would conserve their orbital angular momentum, and only evolve their orbital energy. This would lead to the excitation of their eccentricities, increasing the overall velocity dispersion in the disk, and erasing organized structures like clumps. 

Finally, we note that dust grains may be able to drift out of our simulation domain at $t = 500 P_0$ in models with low values of $\beta$ and high values of ${\rm St}_0$ (e.g., $\beta = 10^{-2}$, ${\rm St}_0 = 10^{-2}$). In those models, the inner drift is so fast that the edge is not able to develop any asymmetry before leaving the domain. This is essentially a numerical issue, and the corresponding markers in the bottom panel of Figure \ref{fig:alpha2} have been removed for clarity. However, it also casts light on the competition between clump formation and inward drift, which plays a critical role in the disk edge dynamics described in Section \ref{sec:recession}.


\subsubsection{The Optical Depth Contribution of Clumps} \label{sec:tauclump}

\begin{figure}[tp]
\centering
\includegraphics[width = 0.45\textwidth]{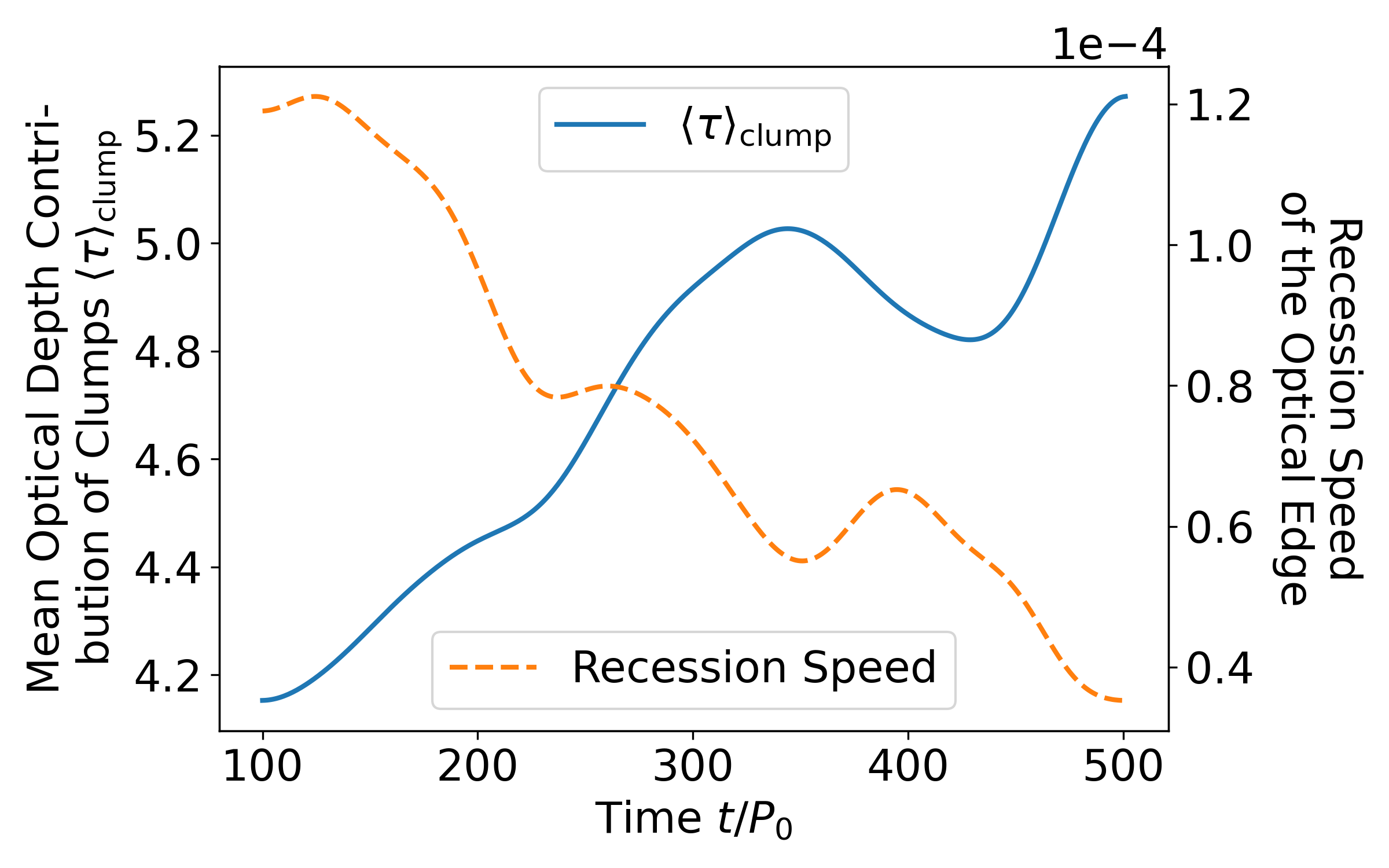}
\figcaption{The time evolution of the mean optical depth contribution of clumps and the recession speed of the optical edge in the fiducial model ($\beta = 10$, ${\rm St}_0 = 10^{-4}$, and $\Sigma_0 = 7.7\times10^{-3}$ g/cm$^2$). Both curves are convolved with a $\sigma_t = 20 P_0$ Gaussian kernel to reduce noise.
\label{fig:dtau}}
\end{figure}

In our models, dust grains only interact through shadowing. This means the optical depths of individual clumps directly determines their influence on dust dynamics. In this section, we investigate the properties and evolution of clumps using their optical depths.

Given a $\tau$ map at a specific time, we first calculate the optical depth contribution ${\rm d}\tau$ of each grid cell. We then seek for locations where ${\rm d}\tau$ has a local maximum in the surrounding $3\times3$ grid and is also greater than a threshold value ${\rm d}\tau_{\rm th1}$. In this way, we sample the overdensities in the disk and rule out false positives due to the Poisson noise. However, this is not sufficient as it also samples the non-isolated overdensities (e.g., the dust wall). So, we then measure the ambient value of ${\rm d}\tau$ around each sampled local maximum by adding up the ${\rm d}\tau$ value in the outer rim of the $5\times5$ grid (16 grid cells), and rule out the ones that have an ambient value larger than another threshold value ${\rm d}\tau_{\rm th2}$. Through trial and error, we find ${\rm d}\tau_{\rm th1} = 0.5$ and ${\rm d}\tau_{\rm th2} = 0.1$ are reasonable threshold values.

Not surprisingly, we find that clumps are optically thick. Their mean optical depth ranges between roughly 2 to 4 at $t = 10~t_{\rm con}$ (Table \ref{tab:clumpsci}). This value is not steady in time --- generally, clumps become more optically thick as time passes. This is likely because the dust wall (which fragments to form clumps) is becoming more massive over time, with grains piling onto it from the inner disk, pushed by radiation pressure, and the outer disk, driven by inward dust drift. The fact that the dust wall is increasing in mass can also be seen indirectly in Figure \ref{fig:dtau}, where it shows that the motion of the dust wall is gradually slowing down as radiation pressure is blocked by more and more grains.

\begin{table}[tp]
\centering
\caption{The mean optical depth contribution of clumps $\tauclump$ with different physical parameters} \label{tab:clumpsci}
\medskip
\begin{tabular*}{0.45\textwidth}{@{\extracolsep{\fill}}ccccc}
\toprule
$\beta$ & St$_0$ & $\Sigma_0$ & $t_{\rm con}$ & $\tauclump$ \\\\[-1.2em]
\midrule
20  & $1.0\times10^{-4}$ & $7.7\times10^{-3}$ & 3.43 $P_0$ & 3.93 \\\\[-1.2em]
10  & $1.0\times10^{-4}$ & $7.7\times10^{-3}$ & 6.85 $P_0$ & 3.73 \\\\[-1.2em]
5   & $1.0\times10^{-4}$ & $7.7\times10^{-3}$ & 13.7 $P_0$ & 2.78 \\\\[-1.2em]
2.5 & $1.0\times10^{-4}$ & $7.7\times10^{-3}$ & 27.4 $P_0$ & 2.39 \\\\[-1.2em]
\midrule
10  & $4.0\times10^{-4}$ & $7.7\times10^{-3}$ & 1.71 $P_0$ & 4.13 \\\\[-1.2em]
10  & $2.0\times10^{-4}$ & $7.7\times10^{-3}$ & 3.43 $P_0$ & 3.86 \\\\[-1.2em]
10  & $1.0\times10^{-4}$ & $7.7\times10^{-3}$ & 6.85 $P_0$ & 3.73 \\\\[-1.2em]
10  & $5.0\times10^{-5}$ & $7.7\times10^{-3}$ & 13.7 $P_0$ & 3.49 \\\\[-1.2em]
10  & $2.5\times10^{-5}$ & $7.7\times10^{-3}$ & 27.4 $P_0$ & 3.14 \\\\[-1.2em]
\midrule
10  & $1.0\times10^{-4}$ & $7.7\times10^{-2}$ & 3.82 $P_0$ & 4.09 \\\\[-1.2em]
10  & $1.0\times10^{-4}$ & $2.5\times10^{-2}$ & 5.18 $P_0$ & 4.20 \\\\[-1.2em]
10  & $1.0\times10^{-4}$ & $7.7\times10^{-3}$ & 6.85 $P_0$ & 3.73 \\\\[-1.2em]
10  & $1.0\times10^{-4}$ & $2.6\times10^{-3}$ & 9.32 $P_0$ & 2.78 \\\\[-1.2em]
\bottomrule
\end{tabular*}
\justify
\tablecomments{The $\tauclump$ values are measured at $t = 10t_{\rm con}$ in all cases, and are convolved with a $\sigma_t = t_{\rm con}$ Gaussian kernel to reduce noise. $\Sigma_0$ is in the unit of g/cm$^2$.}
\end{table}

\begin{figure}[tp]
\centering
\includegraphics[width = 0.45\textwidth]{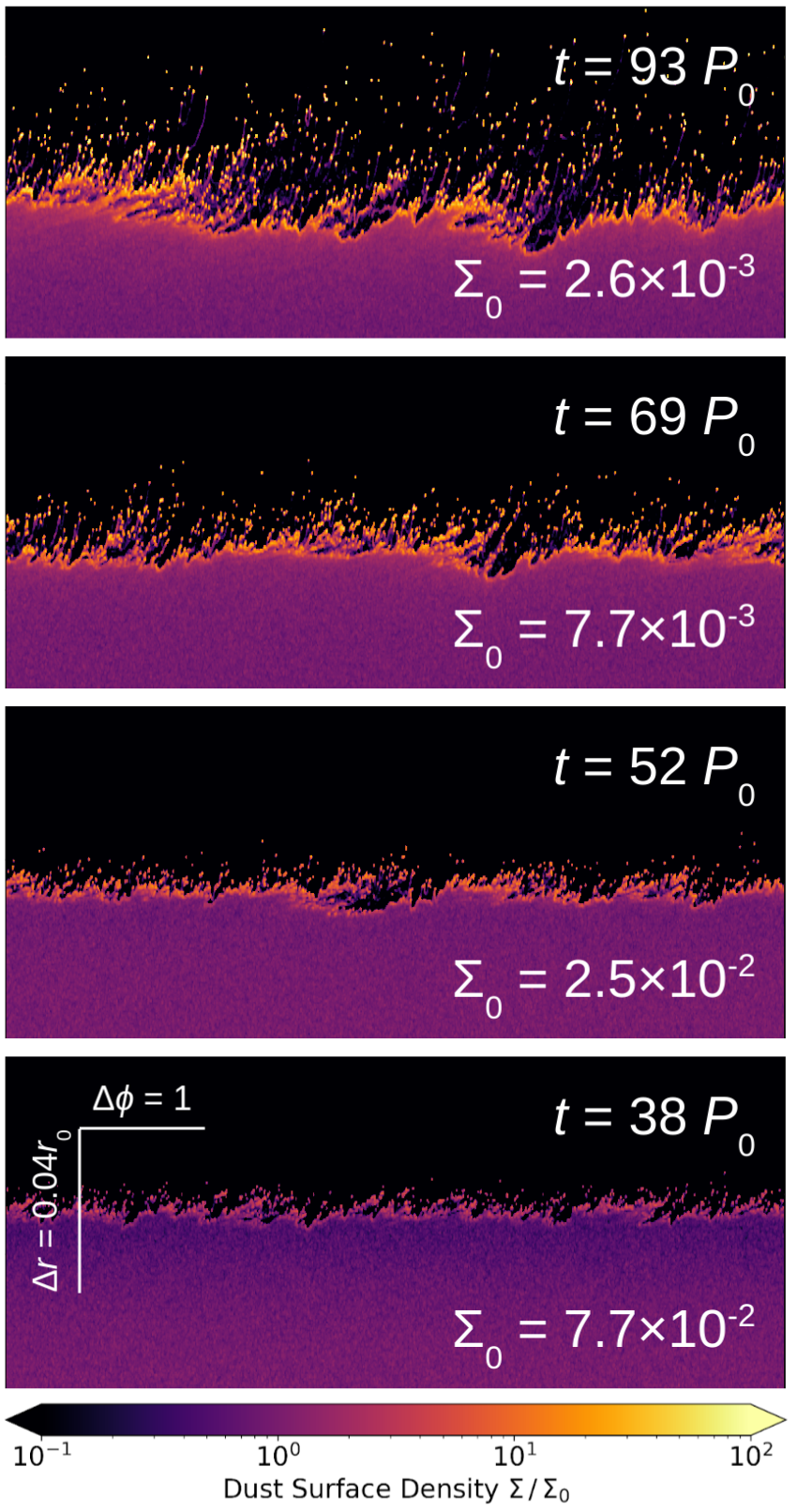}
\figcaption{Snapshots of models with $\beta = 10$, ${\rm St}_0 = 10^{-4}$, but different $\Sigma_0$. Panels are taken at $t = 10t_{\rm con}$ for different models. $\Sigma_0$ is in the unit of g/cm$^2$.
\label{fig:taunorm}}
\end{figure}

Table \ref{tab:clumpsci} also shows that $\tauclump$ is positively correlated to both $\beta$ and ${\rm St}_0$. This is in line with our expectations since clump formation is a competition between radiation pressure, which drives the dust wall's instability, and gas drag, which damps dust motion; and so higher $\beta$ and ${\rm St}_0$ values should make the disk more susceptible to the clumping effect. Meanwhile, we find a rough positive correlation between $\tauclump$ and $\Sigma_0$. To understand this further, Figure \ref{fig:taunorm} shows the morphology of disk edges when $\Sigma_0$ varies. We find that lower $\Sigma_0$ models, even though their clumps are less dense, do have more clumps that are spread over a wider radial extent. Since our density profiles are fixed, a given amount of dust is distributed over a wider radial range in lower $\Sigma_0$ models, at least initially. We therefore speculate that they favor forming smaller clumps because there is a stronger Keplerian shear to overcome in order to form massive clumps.

\begin{table}[tp]
\centering
\caption{The mean optical depth contribution of clumps $\tauclump$ with different numerical parameters} \label{tab:clumpnum}
\medskip
\begin{tabular*}{0.45\textwidth}{@{\extracolsep{\fill}}ccccc}
\toprule
$N_{\rm par}$ & $N_r$ & $N_\phi$ & $t_{\rm con}$ & $\tauclump$ \\\\[-1.2em]
\midrule
$1.0\times10^8$ & 1024 & 1024 & 6.85 $P_0$ & 3.66 \\\\[-1.2em]
$3.3\times10^7$ & 1024 & 1024 & 6.85 $P_0$ & 3.60 \\\\[-1.2em]
$1.0\times10^7$ & 1024 & 1024 & 6.85 $P_0$ & 3.73 \\\\[-1.2em]
$3.3\times10^6$ & 1024 & 1024 & 6.85 $P_0$ & 3.54 \\\\[-1.2em]
$1.0\times10^6$ & 1024 & 1024 & 6.85 $P_0$ & 3.65 \\\\[-1.2em]
\midrule
$2.0\times10^7$ & 2048 & 1024 & 6.85 $P_0$ & 3.86 \\\\[-1.2em]
$1.0\times10^7$ & 1024 & 1024 & 6.85 $P_0$ & 3.73 \\\\[-1.2em]
$5.0\times10^6$ & 512  & 1024 & 6.85 $P_0$ & 3.21 \\\\[-1.2em]
$2.5\times10^6$ & 256  & 1024 & 6.85 $P_0$ & 2.55 \\\\[-1.2em]
\midrule
$2.0\times10^7$ & 1024 & 2048 & 6.85 $P_0$ & 3.44 \\\\[-1.2em]
$1.0\times10^7$ & 1024 & 1024 & 6.85 $P_0$ & 3.73 \\\\[-1.2em]
$5.0\times10^6$ & 1024 & 512  & 6.85 $P_0$ & 3.94 \\\\[-1.2em]
$2.5\times10^6$ & 1024 & 256  & 6.85 $P_0$ & 4.02 \\\\[-1.2em]
\bottomrule
\end{tabular*}
\justify
\tablecomments{The $\tauclump$ values are measured at $t = 10t_{\rm con}$ in all cases, and are convolved with a $\sigma_t = t_{\rm con}$ Gaussian kernel to reduce noise.}
\end{table}

Finally, to check that our results are not sensitive to numerics, we measure $\tauclump$ with different super-particle, radial, and azimuthal resolutions at $t = 10~t_{\rm con}$. The results are shown in Table \ref{tab:clumpnum}. We find good convergence of $\tauclump$ with different values of $N_{\rm par}$ and $N_r$. For $N_\phi$, we reiterate that clumps are eroded (and eventually destroyed) when grains move away from the clump body by one azimuthal grid cell; thus, lower azimuthal resolutions naturally makes it harder for dust grains to leave the clump body, explaining why $\tauclump$ increases when $N_\phi$ decreases in Table \ref{tab:clumpnum}. As $N_\phi$ increases, clumps simply shrink with cell size; it may not be possible to obtain good convergence with respect to $N_\phi$ unless additionally physics, such as some diffusive effects, is included to set a limit to how small clumps can be.


\subsection{Recession of the Disk Edge} \label{sec:recession}

\begin{figure*}[tp]
\centering
\includegraphics[width = 0.9\textwidth]{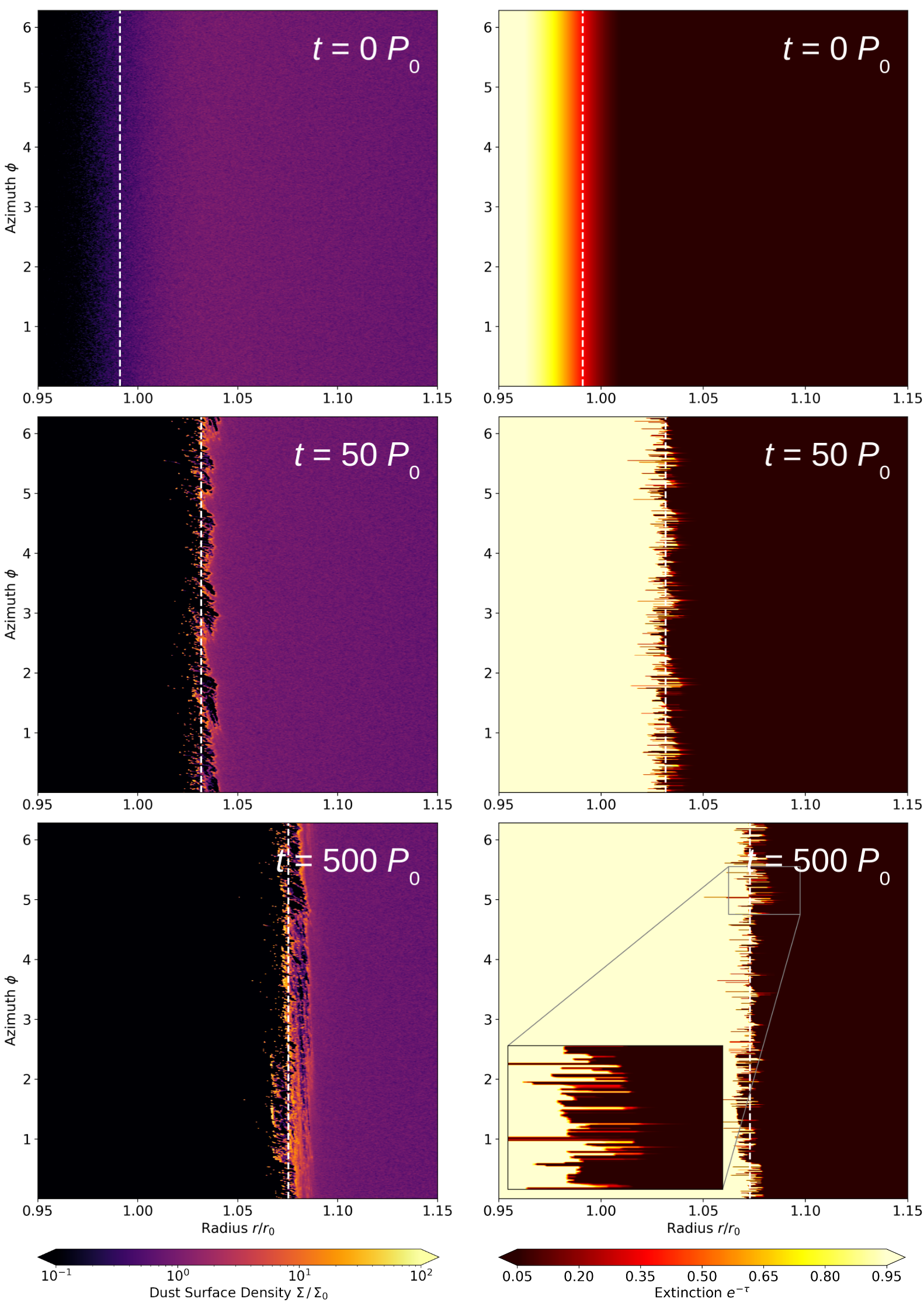}
\figcaption{The time evolution of the dust surface density $\Sigma$ (\textit{left}) and the extinction $e^{-\tau}$ (\textit{right}) in the fiducial model ($\beta = 10$, ${\rm St}_0 = 10^{-4}$, and $\Sigma_0 = 7.7\times10^{-3}$ g/cm$^2$). The dashed line marks the optical edge (where $\tauprime \equiv 1$), separating the optically thin and optically thick regions.
\label{fig:time_evo}}
\end{figure*}

\begin{figure}[tp]
\centering
\includegraphics[width = 0.45\textwidth]{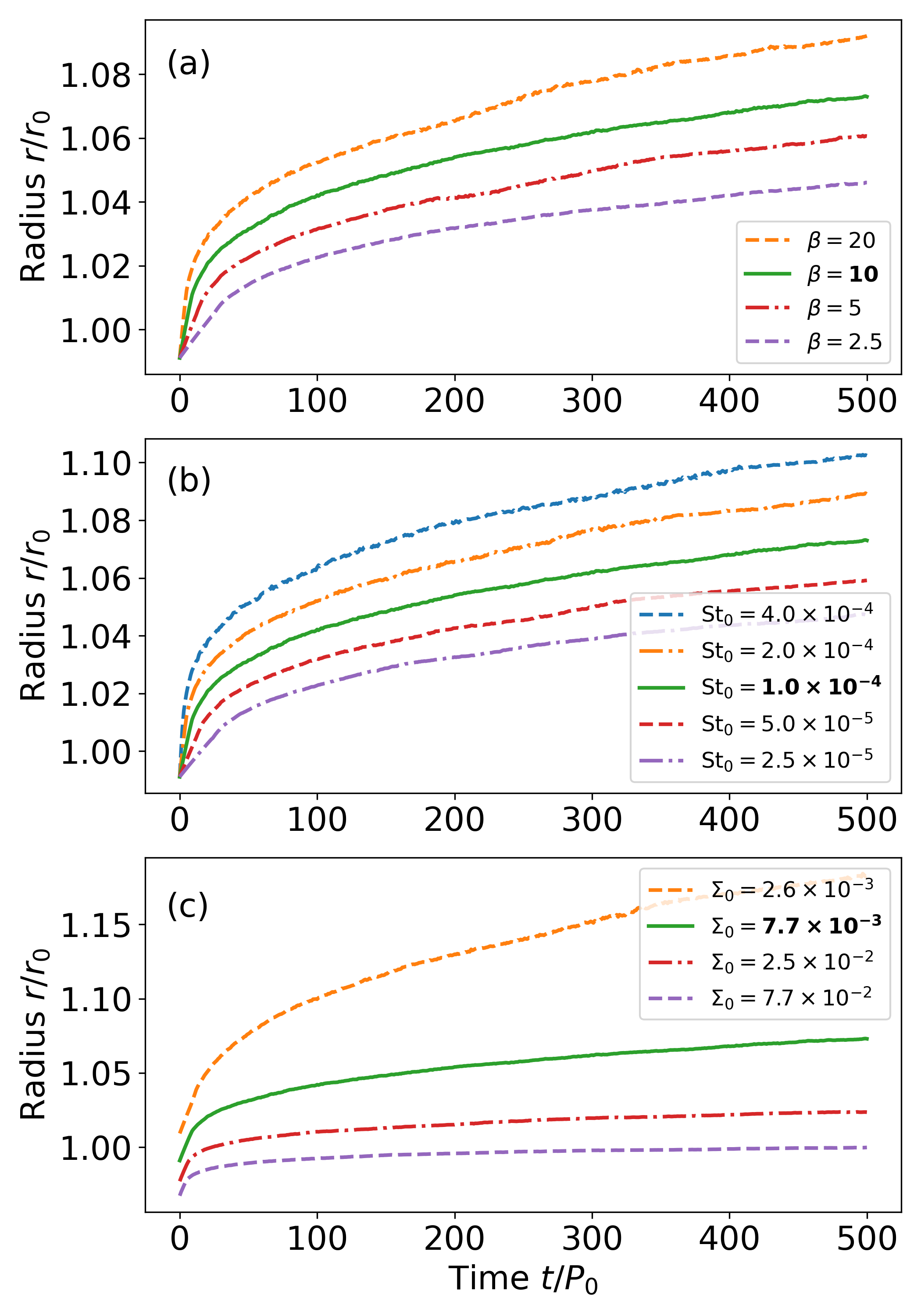}
\figcaption{The time evolution of the optical edge (where $\tauprime \equiv 1$) with different values of \textbf{(a):} ${\rm St}_0$, \textbf{(b):} $\beta$, and \textbf{(c):} $\Sigma_0$. $\Sigma_0$ is in the unit of g/cm$^2$. The fiducial parameters are bold marked in the legend.
\label{fig:sciparamtest}}
\end{figure}

Going back to Figure \ref{fig:asym}, we notice that the radius of the fragmented disk edge (left) is noticeably larger than the axisymmetric edge (right), even though they both used the same initialization. This illustrates another result of the instability, which is that it causes the disk edge to recess, even after the formation of a sharp dust wall. 

The disk edge recesses because radiation leaks through the regions of underdensity at the asymmetric disk edge, exposing the previously shaded disk materials to stronger radiation pressure, and pushing them to higher orbits. This is shown in Figure \ref{fig:time_evo}, where we find multiple ``leaking points'' in the extinction map. Since the leaking points can move azimuthally relative to the disk edge due to the orbital shear, the entire disk is exposed to stronger radiation levels in a time-averaged sense compared with the axisymmetric case in Figure \ref{fig:asym}. This qualitative description is equivalent to the decrease of the overall disk optical depth due to clumping effects (see Figure \ref{fig:alpha1}). Moreover, we note that when the disk edge recesses, the dust density decreases following the geometric factor $r^2$, which leads to a further reduction of the optical depth and promotes recession as a positive feedback.

However, this recession is not guaranteed to last forever. Whether it can be sustained depends on whether it can overcome the inward drift of dust. Dust grains naturally drift inward because the sub-Keplerian rotation of the background gas exerts a headwind on them, gradually removing their angular momentum. As dust grains drift in, they add to the optical depth of the inner edge, making it ``heavier and heavier'' for radiation pressure to push. This competes with the effects that reduce optical depth. If $\tauprime$ increases with time, we expect the edge will eventually stop recessing and start to drift inward. We, in fact, do observe this ``turnover'' in some of our models with ${\rm St}_0 = 10^{-2}$. A similar turnover is also seen in previous 1D studies \citep[e.g.,][]{dominik_accretion_2011}; 1D treatment cannot capture clumping and so the turnover becomes inevitable in those models. We will discuss the turnover further in Section \ref{sec:turnover}. Otherwise, if $\tauprime$ decreases with time (e.g., upper panel of Figure \ref{fig:alpha1}), recession may be sustained.

In the rest of this section, we identify recessing disks in our parameter space, and investigate how $\beta$, ${\rm St}_0$, and $\Sigma_0$ affects their recession speeds. The migration of disk edges are shown in Figure \ref{fig:sciparamtest}. We then compute the recession speed of the optical edge $v_{\rm rec}$ by averaging between $t = 10t_{\rm con}$ and $t = 15t_{\rm con}$ for each model. The results are shown in Table \ref{tab:recess}. 

\begin{table}[tp]
\centering
\caption{The recession speed $v_{\rm rec}$ of the optical edge with different physical parameters}
\label{tab:recess}
\medskip
\begin{tabular*}{0.45\textwidth}{@{\extracolsep{\fill}}cccccc}
\toprule
$\beta$                                     
    & &   20 & \textbf{10}   &    5 &   2.5 \\ \\[-1.2em]
$v_{\rm rec}$ $[10^{-5}r_0\Omega_{\rm K0}]$ 
    & & 5.92 & \textbf{3.06} & 1.31 &  0.75 \\ \\[-1.2em]
\midrule
${\rm St}_0$ $[\times 10^{-4}]$ 
    & 4.0 & 2.0 & \textbf{1.0} & 0.5 & 0.25 \\ \\[-1.2em]
$v_{\rm rec}$ $[10^{-5}r_0\Omega_{\rm K0}]$ 
    & 9.85 & 5.77 & \textbf{3.06} & 1.48 & 0.75 \\ \\[-1.2em]
\midrule
$\Sigma_0$ $[\times10^{-3}\,{\rm g/cm}^2]$                   
    & &  2.6 & \textbf{7.7}  &   25 &    77 \\ \\[-1.2em]
$v_{\rm rec}$ $[10^{-5}r_0\Omega_{\rm K0}]$ 
    & & 5.33 & \textbf{3.06} & 1.76 &  1.65 \\ \\[-1.2em]
\bottomrule
\end{tabular*}
\justify
\tablecomments{The $v_{\rm rec}$ values are averaged from $t = 10t_{\rm con}$ to $t = 15t_{\rm con}$. The fiducial parameters and results are bold marked.}
\end{table}

Our results show that $v_{\rm rec}$ is typically on the order of $10^{-5}\,r_0\Omega_{\rm K0}$, which is considerably fast. Assuming $r_0$ = 1 au, it could result in a cavity that is a few tens of au wide in 1 Myr if it can overcome the accretion flow of the disk. The accretion velocity $v_{\rm acc}$ in a viscous disk model is
\begin{equation}
    v_{\rm acc} = -\frac{3\nu}{2r_0},
\end{equation}
where $\nu$ is the kinematics viscosity. Assuming a constant $\nu = 10^{-6}\, r_0^2\Omega_{\rm K0}$, which is equivalent to $\alpha = 4\times10^{-4}$ at $r_0$ \citep{shakura_reprint_1973}, we have $v_{\rm acc} = -1.5\times10^{-6}\,r_0\Omega_{\rm K0}$. It is certainly within the realms of possibility that this recession of dust can resist and overcome the accretion flow. 

If this recession is sustained over the disk lifetime, we can write:
\begin{equation}
    \frac{{\rm d}r}{{\rm d}t} = 10^{-5}\sqrt{\frac{GM_\star}{r}},
\end{equation}
where $r$ is the location of the disk edge. Given $M_\star$ = $M_\odot$, and a disk age of $\sim$$10^6$ years, we can integrate the above equation and find that $r \approx 20$ au in the end, even if $r$ is negligibly small at first.

We expect stronger radiation pressure (higher $\beta$) and weaker gas drag (higher ${\rm St}_0$) to produce faster $v_{\rm rec}$. The parameter survey shows that $v_{\rm rec}$ is positively correlated to $\beta$ and ${\rm St}_0$ as expected. And the correlations are almost proportional, consistent with the terminal velocity scenario (see Appendix \ref{app:terminal}). We find $v_{\rm rec}$ increases when $\Sigma_0$ decreases, since higher $\Sigma_0$ values naturally make the disk edge heavier for radiation pressure to push outward. Moreover, we find the correlation between $v_{\rm rec}$ and $\Sigma_0$ follows $v_{\rm rec} \propto \Sigma_0^{-0.5}$ when $\Sigma_0$ is not too high (e.g., $\Sigma_0 \leq 2.5\times10^{-2}$ g/cm$^2$). Whether this is a coincidence or there is physics behind it will be explored in the future study.
\section{Discussion} \label{sec:discussion}

\subsection{The Turnover of the Recession} \label{sec:turnover}

The recession of the disk edge may be eventually overcome by a constant influx of dust grains adding to the optical depth of the inner dust wall (see Section \ref{sec:recession}), similar to the disk evolution described in \cite{dominik_accretion_2011}. In this case, the inner edge instability and clump formation can only delay the turnover but not avoid it. Here we discuss the possible conditions for this turnover to occur.

The motion of dust grains at terminal velocity is described in Appendix \ref{app:terminal} (also see \cite{takeuchi_dust_2001}). The radial terminal velocity is given as:
\begin{equation} \label{eq:vrterm}
    v_{\rm r,term} = \frac{{\rm St}}{1 + {\rm St}^2}(\beta e^{-\tau} - \eta).
\end{equation}
Even though $\beta$ is likely orders of magnitude larger than $\eta$ for micron-size grains, the above equation shows that inward drift would still occur when $\tau > \ln{(\beta/\eta)}$. In a simple picture, a constant stream of micron-sized grains flows toward the inner edge from the optically thick portion of the disk, forcing $\tau$ at the inner edge to increase without bound. This is what would ultimately lead to a turnover of the recession.

In reality, not all of the dust grains are fated to end up concentrated at disk edges. They are likely to go through a turbulent evolutionary path, colliding, coagulating, and fragmenting, as they migrate through the disk (e.g., \citealt{laibe_sph_2008, birnstiel_simple_2012, krijt_panoptic_2016, schoonenberg_lagrangian_2018}). Some grains may never reach the inner edge if grain growth is sufficiently fast (e.g., \citealt{okuzumi_rapid_2012, garcia_evolution_2020}). Even at the edge itself, opacity should evolve over time as grain size evolves. Clumps found in our models should accelerate grain size evolution by having higher dust concentration and, hence, higher collision rates. Considering these complications, it is not clear if $\tau$ at the inner edge should always be increasing. If grain growth is able to reduce $\tau$ more rapidly than the addition of small grains through inward drift, we might expect disk recession to continue without a turnover. This clearly demands more sophisticated models than the basic ones presented here, and so whether a turnover might occur, or how long would disk recession last, are questions for future studies.

\subsection{Multi-size Dust Models and Collisions} \label{sec:collision}

We have assumed a single grain size in our proof-of-concept models. Including a distribution of grain sizes is a natural next step, but it would lead to unrealistic results without also considering grain collisions. Revisiting Equation \ref{eq:vrterm}, we note that if $\beta<\eta$ for a given grain size, those grains will always drift inward regardless of the value of $\tau$. For our fiducial parameters, this size is on the order of a millimeter. Assuming a grain size distribution of ${\rm d}n_{\rm s} \propto s^{-3.5} \,{\rm d}s$ \citep{mathis_size_1977}, the optical depth contribution from mm-size grains is $\sim$3\% of that from $\mu$m-size grains. This is a small but significant contribution. Without considering collisions, mm-size grains would drift pass the inner dust wall like ghosts, and shield the entire disk from radiation pressure.

In reality, these larger grains will collide with the inner dust wall. Such collisions tend to favor mass transfer from the larger body to the smaller one \citep{hasegawa_collisional_2021}, which may lead to a net reduction in the local opacity of the disk material, potentially accelerating the disk recession. Collisions, once again, likely plays a critical role.

\subsection{Turbulence} \label{sec:turbulence}

Gas turbulence is another piece of physics that we have not included in our models. In protoplanetary disks, turbulence is often treated as a diffusive or mixing process. Taking this interpretation at face value, turbulence would hamper the generation of dust clumps by diffusing the sharp inner dust wall. However, turbulence is only diffusive on a large scale; on a smaller scale, such as within the thickness of the inner dust wall, turbulence is likely better described as a stochastic perturbation, which makes the wall thinner at some azimuth and thicker elsewhere. In this picture, turbulence may, in fact, seed the density perturbation that leads to clump formation.

Turbulence can also promote grain collisions and facilitate grain size evolution. In our one-size models, the grains are typically so tightly coupled to the gas that, even if there is turbulence, they are unlikely to collide with each other, allowing us to safely ignore this effect. The same cannot be said for models with a distribution of grain sizes. Echoing the previous sections, turbulence, grain collision, and grain size evolution are tightly connected, and may all play important roles in the long-term evolution of the inner disk edge and dust clumps.

\subsection{Outlooks} \label{sec:outlooks}

Dust clumps, in general, are of interest to the topic of planet formation. For instance, self-gravitationally collapsed dust clumps could potentially form planetesimals \citep[e.g.,][]{johansen_rapid_2007, simon_mass_2016}. The clump formation observed in this paper may interact with other clump formation mechanisms, such as streaming instability \citep[e.g.,][]{youdin_streaming_2005, youdin_protoplanetary_2007, johansen_protoplanetary_2007, jacquet_linear_2011}, coagulation instability \citep{tominaga_coagulation_2021}, and some meso-scale instability triggered by the dust feedback in dust rings \citep{huang_meso-scale_2020}. If our clumps do tend to coagulate, a recessing disk edge could leave behind a trail of planetesimals, creating an ideal breeding ground for close-in super-Earths.

How the instability described in this work may operate in a 3D dust disk is another aspect worth investigating. In 2D, the disk is shielded from radiation pressure behind an inner dust wall. But in a flared 3D disk, that wall becomes a shell that envelopes the disk surface. Will there be clumps on the optical surface? How massive can those clumps be? If the dynamics become significantly different from what we have seen in 2D, how will the disk surface react to radiation pressure? These are all important questions to be answered in future studies.
\section{Conclusion} \label{sec:conclusion}

In this paper, we demonstrate a possible mechanism to explain the dust-cleared cavity of transitional disks. We perform simplified, proof-of-concept simulations of dusty disks to show that the inner edge of a dust disk is susceptible to a new radiation-induced instability that amplifies azimuthal asymmetry. We show that:

\begin{itemize}
    \item Dust clumps, which are isolated density features, can form out of the azimuthal density perturbations at the inner cavity edge of the disk.
    \item The clumps make the inner disk edge asymmetric, reduce the effective optical depth of the disk, and lead to rapid outward recession of the disk edge.
\end{itemize}

When the transition of the optical depth is sufficiently sharp at the inner cavity edge of the disk, azimuthal density perturbations can be amplified by the combined effects of radiation pressure, shadowing, and the orbital shear to form clumps (Section \ref{sec:clump}, Figure \ref{fig:clump}). Without internal forces, the clumps in our model are susceptible to destruction due to orbital shear (Section \ref{sec:formation}). The balance between clump formation and destruction maintains a stabilized sharpness of the optical depth transition at the disk edge (Figure \ref{fig:alpha1}). 

The clumps are typically more optically thick than their surroundings (Section \ref{sec:tauclump}). The averaged density of clumps (equivalent to their averaged optical depth contribution $\tauclump$) generally increases with time (Figure \ref{fig:dtau}). The clumps are denser when radiation pressure is stronger (higher $\beta$), or when dust grains are less (but still) coupled to the gas (higher ${\rm St}_0$), or when the disk has more dust (higher $\Sigma_0$; Table \ref{tab:clumpsci}). 

While the clumps can block radiation at specific azimuths, light can leak at other positions and push disk material outward relative to the clumps (Figure \ref{fig:time_evo}). Quantitatively, this can be seen as the clumping effect reducing the effective optical depth of the disk (Figure \ref{fig:alpha1}), which can lead to the disk edge recessing (Section \ref{sec:recession}). In our fiducial model, the recession speed is on the order of $10^{-5}r_0/P_0$, which is sufficiently fast to overcome viscous accretion, if we assume a conventional viscosity of $\alpha = 4\times10^{-4}$. Stronger radiation pressure levels (higher $\beta$), weaker dust-gas coupling (higher ${\rm St}_0$), and lower surface densities (lower $\Sigma_0$) are in favor of faster recession (Table \ref{tab:recess}). 

\section*{Acknowledgement}

We thank Sean Brittain for helpful discussions. We also thank the anonymous referee for constructive suggestions. J.B. acknowledges support from the Education Program for Talented Students of Xi’an Jiaotong University. J.F. gratefully acknowledges support from the Institute for Advanced Study. Numerical calculations are performed on the clusters provided by Compute Canada. 
\clearpage
\bibliography{jiaqing}
\appendix
\section{Terminal Velocities}
\label{app:terminal}

A significant fraction of our parameter employs a small Stokes number ${\rm St}$. The corresponding stopping time is often shorter than the time step of our simulations, and so we would expect the velocities of the super-particles to be closely approximated by their terminal velocities. We solve for these velocities in this appendix.

Without loss of generality, we define the terminal values of the radial speed and specific angular momentum of a super-particle as follows:
\begin{align} \label{eq:terminal}
    v_{\rm r, term} & \equiv u\, v_{\rm K}\\
    l_{\rm term} & \equiv \left(1-L\right) l_{\rm K}\,,
\end{align}
where $u$ and $L$ are dimensionless parameters, and $v_{\rm K}=\sqrt{GM/r}$ and $l_{\rm K}=v_{\rm K} r$ are the Keplerian speed and angular momentum, respectively. At terminal speed, we expect $u$ and $L$ to reach equilibrium, constant values, or, in other words, $\dot{u}=\dot{L}=0$. Without any explicit dependence on time, the radial acceleration can be expressed as $\dot{v}_{\rm r,term}=({\rm d} v_{\rm r,term}/{\rm d}r)v_{\rm r,term}$, and similarly for the torque $\dot{l}_{\rm term}$. This allows us to write the acceleration and torque equations as:
\begin{align}
    \label{eq:vr_term}
    -\frac{1}{2}\frac{v^2_{\rm r, term}}{r} & = -\frac{GM}{r^2}\left(1-\beta e^{-\tau}\right) + \frac{l^2_{\rm term}}{r^3} - \frac{v_{\rm r, term}}{{\rm St}\, \Omega_{\rm K}^{-1}}\\
    \label{eq:l_term}
    \frac{1}{2}\frac{l_{\rm term}v_{\rm r, term}}{r} & = -\frac{l_{\rm term}-l_{\rm gas}}{{\rm St}\, \Omega_{\rm K}^{-1}}\,.
\end{align}
Dividing Equation \ref{eq:vr_term} by $v^2_{\rm K}/r$ and \ref{eq:l_term} by $v^2_{\rm K}$, we obtain the dimensionless equations:
\begin{align}
    -\frac{u^2}{2} & = -1 + \beta_{\rm eff} + \left(1-L\right)^2 - \frac{u}{{\rm St}}\\
    \frac{u\left(1-L\right)}{2} & = \frac{-1 + L + f_{\rm gas}}{{\rm St}} \,,
\end{align}
where we have, for convenience, defined $\beta_{\rm eff} \equiv \beta e^{-\tau}$ and $l_{\rm gas} \equiv f_{\rm gas}l_{\rm K}$ such that $f_{\rm gas}\equiv\sqrt{1-\eta}$. Solving for $u$ and $L$, we obtain:
\begin{align}
    \label{eq:u_exact}
    u & = \frac{1 - \sqrt{1 - 2 {\rm St} u_0}}{{\rm St}}\\
    \label{eq:L_exact}
    L & = \frac{1 - f_{\rm gas} + \frac{{\rm St}}{2} u}{1 + \frac{{\rm St}}{2}u} \,,
\end{align}
where $u_0 = {\rm St}\left(\beta_{\rm eff} -2L + L^2\right)$. This set of equations has no simple analytic solution, but can be solved through iterations to any desired precision. A good starting point is to assume ${\rm St}\,u_0\ll1$, which allows us to write:
\begin{equation}
    u = u_0 + O({\rm St}\,u_0^2) \,,
\end{equation}
and
\begin{equation}
    L = 1 - f_{\rm gas} + f \frac{{\rm St}}{2} u_0 +  O(({\rm St}\,u_0)^2)\,.
\end{equation}
Dropping terms of order $({\rm St}\,u_0)^2$ and smaller, we can solve for $L$:
\begin{equation}
    \label{eq:L_approx}
    L \approx \frac{1 + {\rm St}^2 f_{\rm gas}}{{\rm St}^2 f_{\rm gas}}\left(1-\sqrt{1 - 2  \frac{{\rm St}^2 f_{\rm gas}\left(1-f_{\rm gas} + {\rm St}^2 f_{\rm gas} \beta_{\rm eff}/2\right)}{(1+{\rm St}^2 f_{\rm gas})^2}}\right)\,,
\end{equation}
and then obtain $u$ by plugging $L$ back into Equation \ref{eq:u_exact}. Using this approximation as the initial guess and applying {\rm St}effensen's method for iterative root finding, we can solve Equations \ref{eq:u_exact} and \ref{eq:L_exact} to double precision numerical accuracy in typically one or two iterations.

While we use Equations \ref{eq:u_exact}, \ref{eq:L_exact}, and \ref{eq:L_approx} to obtain numerically accurate solutions, it is helpful to also simplify the equations further to gain some intuition into the problem. Taking $\eta\ll1$ and keeping only leading order terms in $\eta$, $u_0$, and $L$, we get:
\begin{align}
    u & \approx \frac{{\rm St}}{1+{\rm St}^2}\left(\beta_{\rm eff} - \eta\right) \\
    L & \approx \frac{\eta + {\rm St}^2 \beta_{\rm eff}}{2\left(1+{\rm St}^2\right)}\,.
\end{align}
We use these approximated solutions to inform our analysis in Section \ref{sec:results}.


\end{document}